\documentclass[%
 reprint,
 amsmath,amssymb,
 aps,floatfix
]{revtex4-2}

\usepackage[usenames,dvipsnames]{xcolor}

\let\vec\mathbf
\def\mb{\mathbf}

\usepackage{soul} 
\usepackage{graphicx}
\usepackage{dcolumn}
\usepackage{bm}
\usepackage[colorlinks=true,linkcolor=blue,citecolor=blue,urlcolor=blue]{hyperref}
\usepackage{float}
\newcommand*{\affaddr}[1]{#1} 
\newcommand*{\affmark}[1][*]{\boldsymbol{\textsuperscript{#1}}}
\newcommand{\tens}[1]{%
  \mathbin{\mathop{\otimes}\displaylimits_{#1}}%
}

\def\be{\begin{equation}}
\def\ee{\end{equation}}
\begin{document}

\title{Phase transition dimensionality crossover from two to three dimensions\\
in a trapped ultracold atomic Bose gas}
\date{\today}
\email{n.a.l.keepfer1@newcastle.ac.uk}
\author{%
N. A. Keepfer\affmark[1,2], I.-K. Liu\affmark[1], F. Dalfovo\affmark[2], N. P. Proukakis\affmark[1]\\
\affaddr{\affmark[1]Joint Quantum Centre (JQC) Durham-Newcastle, School of Mathematics, Statistics and Physics, Newcastle University,
Newcastle upon Tyne, NE1 7RU, United Kingdom}\\
\affaddr{\affmark[2]INO-CNR BEC Center and Dipartimento di Fisica,
Universit\`a di Trento, via Sommarive 14, I-38123 Trento, Italy}\\
}

\begin{abstract}
The equilibrium properties of a weakly interacting atomic Bose gas across the Berezinskii-Kosterlitz-Thouless (BKT) and Bose-Einstein condensation (BEC) phase transitions are numerically investigated through a dimensionality crossover from two to three dimensions. The crossover is realised by confining the gas in an experimentally feasible hybridised trap which provides homogeneity along the planar $xy$-directions through a box potential in tandem with a harmonic transverse potential along the transverse $z$-direction. The dimensionality is modified by varying the frequency of the harmonic trap from tight to loose transverse trapping. Our findings, based on a stochastic (projected) Gross-Pitaevskii equation, showcase a continuous shift in the character of the phase transition from BKT to BEC, and a monotonic increase of the identified critical temperature as a function of dimensionality, with the strongest variation exhibited for small chemical potential values up to approximately twice the transverse confining potential.
\end{abstract}

\maketitle

\section{\label{sec:Intro}Introduction\protect}

The behaviour of Bose-Einstein condensates (BECs) \cite{Pethick08Bose,Pitaevskii16Bose,Dalfovo99Theory} around the critical transition temperature is a well researched topic in the field of quantum gases. In a three-dimensional (3D) trapping geometry, a dilute Bose gas condenses into the lowest available energy state, manifesting a transition to a superfluid phase. Instead, in a strictly two-dimensional (2D) trapping geometry, the formation of a condensate is precluded by the Mermin-Hohenberg-Wagner theorem \cite{Mermin66absence,Hohenberg67Existence}. In this instance, the Berezinskii-Kosterlitz-Thouless (BKT) transition \cite{Berezinskii72Destruction,Kosterlitz73Ordering} occurs in its place; here, bound vortex pairs provide a topological ordering and superfluidity at low enough temperatures. Although both regimes have been well-characterized both theoretically and experimentally in ultracold atomic gases, 
an interesting emerging question concerns the dependence of the phase transition characteristics on dimensionality, between these two paradigms. Whilst dimensionality crossovers across the 2D-3D regime have been previously considered both theoretically 
\cite{Mullin97Bose,Stoof02Phase,Stoof02Low,Stoof02Erratum,Stoof03Dimensional,Petrov00Bose,Petrov04Low,Van2002Dilute,Bisset09Transition},
and experimentally~\cite{Ketterle01Realization,Hadzibabic06Berezinskii,Kruger07Critical,Hadzibabic08The,Hadzibabic11Two,Chomaz15Emergence,Fletcher15Connecting},
a systematic theoretical analysis of the effects of dimensionality between the BKT and BEC phase transitions in a dilute Bose gas is still lacking.

In this work we characterise the 2D-3D dimensionality crossover using as a reference case an experimentally viable geometry, such as the setting employed by the group of J.~Dalibard \cite{Ville18Sound}. The gas is confined in a rectangular box potential in the $xy$-plane and, at the same time, in a harmonic potential along the transverse $z$-direction. We perform numerical simulations for various values of the transverse frequency on a wide range, from very tight to loose trapping. Our simulations are based on a stochastic projected Gross-Pitaevskii equation (SPGPE)
\cite{Stoof01Dynamics,Duine01Stochastic,Proukakis03Coherence,Gardiner03The,Blakie08Dynamics,Bradley08Bose,Weiler08Spontaneous,Proukakis08Finite,Proukakis09The,Proukakis13Quantum,Berloff14Modeling}.
For each set of parameters, we average over many SPGPE trajectories in order to extract relevant statistical properties of the gas, such as the mean occupation of the lowest energy states, the Binder cumulant, 
the quasi-condensate density and the condensate fraction. All these quantities are found to interpolate smoothly between the expected behaviours in the 2D and 3D limits. Their study allows us to determine the range of parameters over which this crossover occurs,
and specifically the dependence of the critical temperature on dimensionality; the latter reveals a monotonically increasing trend from 2D to 3D, but with most variation manifesting itself over rather tight traps for which the chemical potential is less than about twice the transverse oscillator energy.
We believe such identification will prove useful for a range of finite temperature experiments conducted in rather (but not necessarily very) tightly-confined quasi-2D traps. 

This paper is structured as follows:
In Sec.~\ref{sec:Theory} we present the system under study and our theoretical method and numerical scheme. In Sec.~\ref{sec:results}, we first highlight the parameters used to characterize the phase transition (Sec.~\ref{sec:Parameters}), before demonstrating the consistency of our results with known 2D (Sec.~\ref{sec:2DLimit}) and 3D limits (Sec.~\ref{sec:3DLimit}).
Equipped with such tools, Sec.~\ref{sec:Quasi2D} presents our main results associated with the entire dimensionality crossover, with key findings summarized in Sec.~\ref{sec:conclusion}, and further technical details contained in the Appendices.

\section{\label{sec:Theory}Theoretical Method}

In order to study system characteristics and identify the transition temperature we utilise the SPGPE~\cite{Stoof01Dynamics,Duine01Stochastic,Proukakis03Coherence,Gardiner03The,Blakie08Dynamics,Bradley08Bose,Weiler08Spontaneous,Proukakis08Finite,Proukakis09The,Proukakis13Quantum,Berloff14Modeling}. This model includes fluctuations and spontaneous processes and has been previously successfully applied through the transition temperature to model both equilibrium~\cite{Cockburn12Ab,Comaron19Quench} and dynamical~\cite{Weiler08Spontaneous,Damski10Soliton,Su13Kibble,Rooney13Persistent,Kobayashi16Thermal,Kobayashi16Quench,Liu18Dynamical,Ota18Collisionless,Comaron19Quench,Liu20Kibble} properties in both 2D and 3D settings, thus allowing an accurate numerical  determination of the critical temperature across the full crossover probed in this work. Related classical field studies of the BKT phase transition have been conducted in harmonic traps~\cite{Simula06Thermal,Simula08Superfluidity,Bisset09Quasi,Bezett09Critical,Mathey17Dynamic} and box-like geometries~\cite{Foster10Vortex,Karl17Strongly,Gawryluk19Signatures}.

Within the SPGPE model, the atomic gas is separated into two parts: a number of highly-occupied low-energy modes described by the classical Langevin field, or simply c-field,  $\Psi(\mathbf{r}, t)$ up to a cutoff energy $\epsilon_{\mathrm{cut}}$, and a thermal reservoir of incoherent states above the cutoff, assumed to be in constant equilibrium with the classical field.
The dynamics of the c-field are governed by 
\cite{Blakie08Dynamics,Proukakis08Finite,Proukakis13Quantum}
 \begin{equation}
      \begin{aligned}
i \hbar \frac{\partial \Psi(\mathbf{r}, t)}{\partial t}=&\mathcal{P}\bigg{\{}(1-i \gamma)\bigg(-\frac{\hbar^{2} \nabla^{2}}{2 m}+V(\mathbf{r})\\& +g|\Psi(\mathbf{r}, t)|^{2}-\mu\bigg)
\Psi(\mathbf{r}, t)+\eta(\mathbf{r}, t)\bigg{\}}.
\end{aligned}
\label{eqn:SPGPE}
 \end{equation}
Here $\gamma$ is a dimensionless parameter controlling the rate of relaxation of the c-field modes to the equilibrium configuration; the value of $\gamma$ is related to the strength of noise correlations $\eta$, describing the interaction of $\Psi$ with the high-lying thermal reservoir,
 \begin{equation}
     \left\langle\eta^{*}(\mathbf{r}, t) \eta\left(\mathbf{r}^{\prime}, t^{\prime}\right)\right\rangle=2 \gamma k_{B} T  \delta_c\left(\mathbf{r}-\mathbf{r}^{\prime}\right) \delta\left(\mathbf{t}-\mathbf{t}^{\prime}\right) ,
\end{equation}
where the notation $\langle\ldots\rangle$ describes averaging over different noise realisations and  $\delta_c(\mathbf{r}-\mathbf{r}')\equiv\sum_{\epsilon\leq\epsilon_\mathrm{cut}}\varphi_\epsilon^\ast(\mathbf{r})\varphi_\epsilon(\mathbf{r}')$ for a basis $\varphi_\epsilon$ labeled by energy $\epsilon$ \cite{Blakie08Dynamics}. 
The energy cutoff is identified with the expression $\epsilon_{\mathrm{cut}}=\mu+k_{B} T \log 2$, which gives a mean occupation number of order $1$ for the states of an ideal Bose gas near the cutoff~\cite{Rooney10Decay}.
The symbols $k_B$, $T$ and $\mu$ refer to the the Boltzmann constant, temperature of the thermal reservoir and chemical potential respectively; 
$m$ is the atomic species mass and $g=4\pi\hbar^2a_s/m$ is the inter-atomic interaction strength, with $a_s$ as the s-wave scattering length, and 
$\mathcal{P}$ is a projector to constrain the dynamics within the c-field region.
The atoms above the cutoff do still contribute to the system through the total atomic density, whose equilibrium value is self-consistently evaluated as discussed below.

To investigate the dimensionality crossover in the most realistic manner, we emulate a previous experimental work with a gas of ${ }^{87} \mathrm{Rb}$ atoms trapped in a box-harmonic hybrid potential (see Sec.~2.3 of \cite{Ville18Quantum}) given by
 \begin{equation}
     V(\mathbf{r})= V_{\rm box}(x,y)+\frac{1}{2}m\left( \frac{\omega_{\rm ref}}{\Lambda}\right)^2 z^2,
     \label{eqn:potential}
 \end{equation}
where $\omega_{\rm ref}=2\pi\times 4.59\;$kHz as in Ref.~\cite{Ville18Sound}. Here, $V_{\rm box}(x,y)$ is zero within a hard-walled rectangular planar box of size $L_x=38\mu$m, $L_y=30\mu$m, and very large outside. 
To control the tightness of the harmonic confinement, we have introduced a dimensionless parameter $\Lambda$, such that the transverse trapping frequency is $\omega_z=\omega_{\rm ref}/\Lambda$ and one can define a typical transverse length $\ell_z = \sqrt{\hbar/m\omega_z} = \sqrt{\Lambda} \, l_{\rm ref}$, where in our case $l_{\rm ref}=\sqrt{\hbar/m\omega_{\rm ref}}= 0.1592 \ \mu$m. Finally, the scattering length is $a_s = 5.09$~nm, which implies $g= (0.064\, \mu{\rm m}) \hbar^2/m$.

To solve the SPGPE, we expand the c-field wave function in a hybrid basis, $\Psi(\mathbf{r},t)=\sum_{\epsilon \leq\epsilon_\mathrm{cut}}A_{\epsilon}(t)\varphi_{\epsilon}(\mathbf{r})$, where $\varphi_{\epsilon}(\mathbf{r})$ is the product of plane waves along $x$ and $y$ and the eigenfunctions of the harmonic oscillator along the $z$ direction (See Appendix~\ref{App:Numerics} for numerical details). Due to the finite size of the box, the momenta along $x$ and $y$ are discrete and the energy of the single particle states of the basis can be labelled by three integer quantum numbers. The solution of the SPGPE provides the amplitudes $A_{\epsilon}(t)$ and hence the c-field $\Psi(\mathbf{r},t)$. To this aim we use the fast Fourier and Hermite transformations, where the Hermite transformation can be accurately calculated with the implementation of inhomogeneous Hermite grid~\cite{Blakie08Dynamics,Blakie08Numeric}. From the solution we can calculate the density in the c-field, $n_c(\mathbf{r},t)= |\Psi(\mathbf{r},t)|^2$ and the corresponding atom number, $N_c=\int d{\textbf{r}} |\Psi|^2$. 

The number the atoms in the incoherent states above the energy cutoff, $N_\mathcal{I}$, as well as their density, $n_\mathcal{I}$, can be estimated assuming that those states are single particle states of an ideal gas in the same hybrid trap and that their mean occupation number at equilibrium is determined by the Bose-Einstein distribution, so that $N_\mathcal{I} = \sum_{\epsilon>\epsilon_\mathrm{cut}} 1/\{\exp[(\epsilon-\mu)/k_BT]-1\}$, where $T$ and $\mu$ are the same of the c-field solution of the SPGPE (see Appendix \ref{App:density} for details). 
 \begin{figure*}[!ht]
    \begin{center}
        \includegraphics[width=0.95\linewidth, clip]{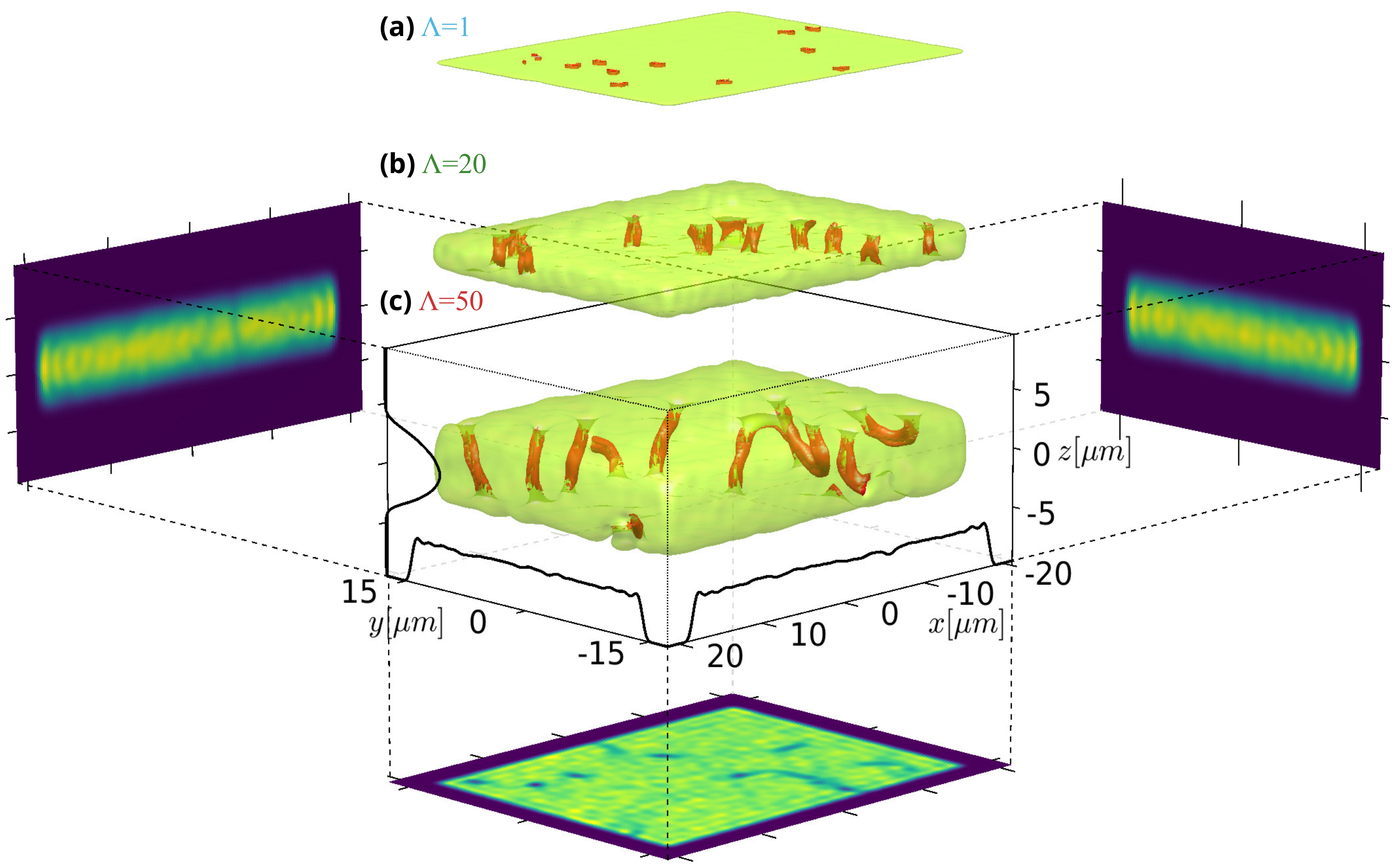}
    \end{center}
    \caption{\protect Examples of isosurface rendering (green) of c-field density, $|\Psi|^2$, during a quench from zero-field conditions to equilibrium for $\Lambda={1,20,50}$ as labelled by (a), (b) and (c) respectively. The temperature of the gas in all cases is $20\mathrm{nK}$, which is well below the transition temperature where the system is superfluid. In red, we plot an isosurface of high velocity regions, indicating vortex structures. In the $\Lambda=50$ case, integrated line density profiles are overlayed for each axis. The image at the bottom is the column density, i.e., the density integrated along $z$, for the case of $\Lambda=50$.  The images on the left and right are the column densities in the planar directions.}
    \label{fig:Figure1}
\end{figure*}

Our aim is to compare configurations with similar average density but different dimensionality, as this facilitates the most direct way to compare different regimes across the dimensionality crossover. However, the SPGPE does not allow one to directly fix the total atom number $N=N_c+N_\mathcal{I}$ as an input. One instead has to impose upon the thermal reservoir, and consequently the Bose gas, an input chemical potential. A simple choice corresponds to use $\mu$ according to
\begin{equation}
    \mu  = \mu_{\textrm{2D}} + \frac{\hbar\omega_{z}}{2} = \mu_{\textrm{2D}} 
    + \frac{\hbar\omega_{\rm ref}}{2 \Lambda},
    \label{eq:inputmu}
\end{equation}
where $\mu_{\textrm{2D}}$ is a constant, independent of $\Lambda$.  When $\Lambda$ increases, this choice of $\mu$ implies an increase of the number of atoms in the trap: typical c-field atom numbers for a given ensemble range between $N_c\sim10^3$ for $\Lambda=1$ to $N_c\sim10^6$ for $\Lambda=50$. However, as we will see later, the central density remains of the same order in the whole range of $\Lambda$, decreasing by about $30\%$ only and approaching the constant value $\mu/g$ for $T=0$ and large $\Lambda$. The results presented in the next sections are obtained by using the chemical potential (\ref{eq:inputmu}) with $\mu_{\textrm{2D}}=(4.64  \mu\textrm{m}^{-2}) \hbar^2/m$, which ensures that the density and the chemical potential for $\Lambda=1$ reduce to the experimental values of Ref.~\cite{Ville18Sound}. In Section~III.D, we will also discuss the results obtained with a chemical potential having a different dependence on $\Lambda$, in order to show that main qualitative features of the dimensional crossover remain unchanged.   

To prepare equilibrium configurations for each value of $\Lambda$ and temperature, we evolve the system dynamically from a zero-field initial condition for a time $t \sim 100 \tau_\gamma$, where $\tau_\gamma=\hbar/\mu\gamma$.
Throughout this work we employ a value $\gamma=0.05$ as a reasonable estimate for our system, noting that similar values have been used in previous SPGPE simulations \cite{Ota18Collisionless,Roy21Finite,Comaron19Quench,Larcher18Dynamical}; nonetheless, we stress that the precise value of $\gamma$ is not relevant for the present work, as it determines the rate at which the system relaxes to equilibrium, but has little effect on the properties of the system once equilibrium is reached. To obtain distinct/independent realizations at fixed $\Lambda$ and $T$ for our subsequent stochastic analysis, we then propagate such equilibrium solution further in time with Eq.~(\ref{eqn:SPGPE}), sampling an additional realisation every $\sim 10 \tau_\gamma$. Such a procedure is justified under the ergodic principle, within which the time average over the evolution of a single system at equilibrium is indiscernible from the ensemble average over many
different systems \cite{Davis02Simulations,Reichl99Modern}, and provides a significant numerical speed-up to creating distinct equilibrium stochastic realizations through full dynamical equilibration based on a different initial noisy zero-field configuration. For a given $\Lambda$, we prepare between $\mathcal{N}=50-100$ realizations for each probed temperature, and probe each distinct temperature, $T$, in $10\mathrm{nK}$ increments from $10\mathrm{nk}$ up to $300\mathrm{nk}$. This procedure  gives a thermal resolution of $\pm 5\mathrm{nK}$ in identifying critical behaviour of the phase transition across the range of $\Lambda$ we consider. 

Before discussing the equilibrium properties, we first visualise in Fig.~\ref{fig:Figure1} a few snapshots of single SPGPE trajectories for three different values of $\Lambda$ during the preparatory stage of the dynamical equilibration process. In each case, the system evolves starting from a zero-field condition at a temperature $T$ just below the transition temperature and each snapshot is taken at some instant during the equilibration dynamics. In such a dynamical process, one may easily observe quantum vortex structures forming spontaneously during the growth~\cite{Weiler08Spontaneous,Proukakis09The,Damski10Soliton,Liu18Dynamical,Adolfo14Universality}.
The figure shows that, by increasing $\Lambda$, vortices change from point-like to filament-like defects, clearly signposting a transition in dimensionality, which roughly occurs when the vortex core size (i.e., the healing length) is of the same order of the transverse size of the atomic cloud.  

\section{\label{sec:results}Results}

In this section, we discuss the physical parameters probed to perform our analysis. We then consider the two limiting regimes of our system, namely the 2D limit ($\Lambda = 1$) and the 3D limit (asymptotically approached at high values of $\Lambda$, with such role played by our chosen value of $\Lambda = 50$), demonstrating agreement of our results with predictions in such limiting regimes. This then leads to the main part of this manuscript, namely the crossover behaviour.

\subsection{\label{sec:Parameters}Parameters characterizing the phase transition }

To locate the critical region of the phase transition, we use a set of different relevant equilibrium quantities, in close analogy to earlier works~\cite{Bisset09Quasi,Bezett09Critical,Foster10Vortex,Kobayashi16Quench,Comaron19Quench,Liu20Kibble}.

Firstly, to identify the existence of a condensate in the system, we use a standard procedure \cite{Blakie08Dynamics} based on extracting atom numbers, $N_i$, corresponding to the $i$-th mode through Penrose-Onsager diagonalisation \cite{Penrose56Bose} of the one-body density matrix $\rho\left(\mathbf{r}, \mathbf{r}^{\prime}\right)=\left\langle\Psi^{*}(\mathbf{r})\tens{} \Psi\left(\mathbf{r}^{\prime}\right)\right\rangle_{\mathcal{N}}$, where $\langle\ldots\rangle_\mathcal{N}$ denotes an averaging over $\mathcal{N}$ stochastic realisations. Eigenvalues extracted in this manner allow, for $i=0$, the reconstruction of the corresponding mode, $\psi_0(\vec{r})$, with the largest eigenvalue, which we henceforth identify as the condensate mode.  Beyond identifying the condensate fraction $N_0/N$, we can also extract the ratio, $N_1/N_0$, of atoms in the second lowest to lowest (condensate) mode. 

Condensation refers to a state with stable coherent phase and a single macroscopically occupied mode, arising when both density and phase fluctuations are suppressed. 
While this is highly-relevant in 3D systems (below the regime of critical fluctuations), lower dimensional systems feature pronounced phase fluctuations even when density fluctuations are suppressed~\cite{Stoof02Low,Stoof02Erratum,Petrov04Low}, leading to a state of quasicondensation, which spans multiple microscopically occupied modes \cite{Kagan92Kinetics,Petrov01Phase}.
The latter is numerically determined via~\cite{Prokof01Critical,Prokof02Two}
\begin{equation}
     N_{\rm QC}=\sqrt{2\left\langle |A_{0}|^2\right\rangle_\mathcal{N}^2-\left\langle |A_{0}|^4\right\rangle_\mathcal{N}} \ ,
\end{equation}
where $A_0 = \int d\mathbf{r}~\psi_{0}^\ast(\mb{r})\Psi(\mb{r})$.
We can thus introduce a measure of the difference between quasicondensation and Bose-Einstein condensation through the parameter
\begin{equation}
    \zeta = \frac{N_{\mathrm{QC}}-N_{0}}{N_{\mathrm{QC}}}
\end{equation}
which should reveal a noticeable difference in behaviour in the 2D, where BEC is precluded, and 3D cases. 

An important quantity commonly used to characterise the critical region is the 
Binder cumulant \cite{Binder81Finite,Davis03Microcanonical,Liu20Kibble,Foster10Vortex,Kobayashi16Thermal,Kobayashi16Quench,Comaron19Quench}, defined by
\begin{equation}
    C_B=\frac{\left\langle |A_{0}|^4\right\rangle_\mathcal{N}}{\left\langle |A_0|^2\right\rangle^2_\mathcal{N}} \ ,
    \label{eqn:binder}
\end{equation}
which is known to display critical behaviour across the phase transition, yielding -- in the limit of infinitely large 3D boxes -- a step-like behaviour from $1$ (fully coherent system) to $2$
(pure thermal state), with such transition being smoothed in finite systems, or due to the presence of harmonic confinement~\cite{Foster10Vortex,Comaron19Quench,Liu20Kibble}.

Such quantities are calculated below both in the 2D and 3D limiting cases -- for ensuring consistency with established results -- and in the entire crossover region.

\subsection{\label{sec:2DLimit}2D limit: BKT transition }

The theory of Berezinskii, Kosterlitz and Thouless (BKT) stipulates that a topologically-induced phase transition may occur in the 2D Bose gas, below some critical temperature. At high temperatures, above the transition, there is a proliferation of free point-like vortices, which apply phase defects to the fluid of an integer value of $2\pi$. This proliferation of free vortices inhibits phase rigidity within the system, leading to a fluctuating phase and vanishingly small correlation lengths. As the temperature is lowered, vortices bind together into dipole pairs, bearing little influence on the coherence properties of the gas yet allowing the phase to stabilise and become coherent over a larger correlation length. This topological ordering of bound vortex pairs provides a mechanism for the BKT phase transition. 

Previous works \cite{Fisher88SDilute,Prokof01Critical} have developed a thermodynamic theory by determining the equation of state for an interacting planar 2D Bose gas, leading to the critical expression in the thermodynamic limit described by
\begin{equation}
    \left.\frac{\mu_{\mathrm{2D}}}{k_{\mathrm{B}} T}\right|_{\mathrm{BKT}} \approx \frac{mg_{\mathrm{2D}}}{\pi \hbar^2} \ln \left(\frac{C_\mu}{\tilde{g}_{\mathrm{2D}}}\right),
\label{eq:tbkt}
\end{equation}
where $\mu_{\mathrm{2D}}$ is the effective chemical potential for the 2D system, as in Eq.~(\ref{eq:inputmu}). The quantity $\tilde{g}_{\mathrm{2D}}= (m/\hbar^2)\, g_{\rm 2D} = \sqrt{8\pi} a_{s}/l_{\rm ref} =0.16$ is a dimensionless interaction strength in the 2D limit chosen to once again match the value used in Ref.~\cite{Ville18Sound}, while $C_\mu = 13.2$ \cite{Prokof02Two}.

\begin{figure}[!ht]
    \centering
    \includegraphics[width=0.95\linewidth,keepaspectratio]{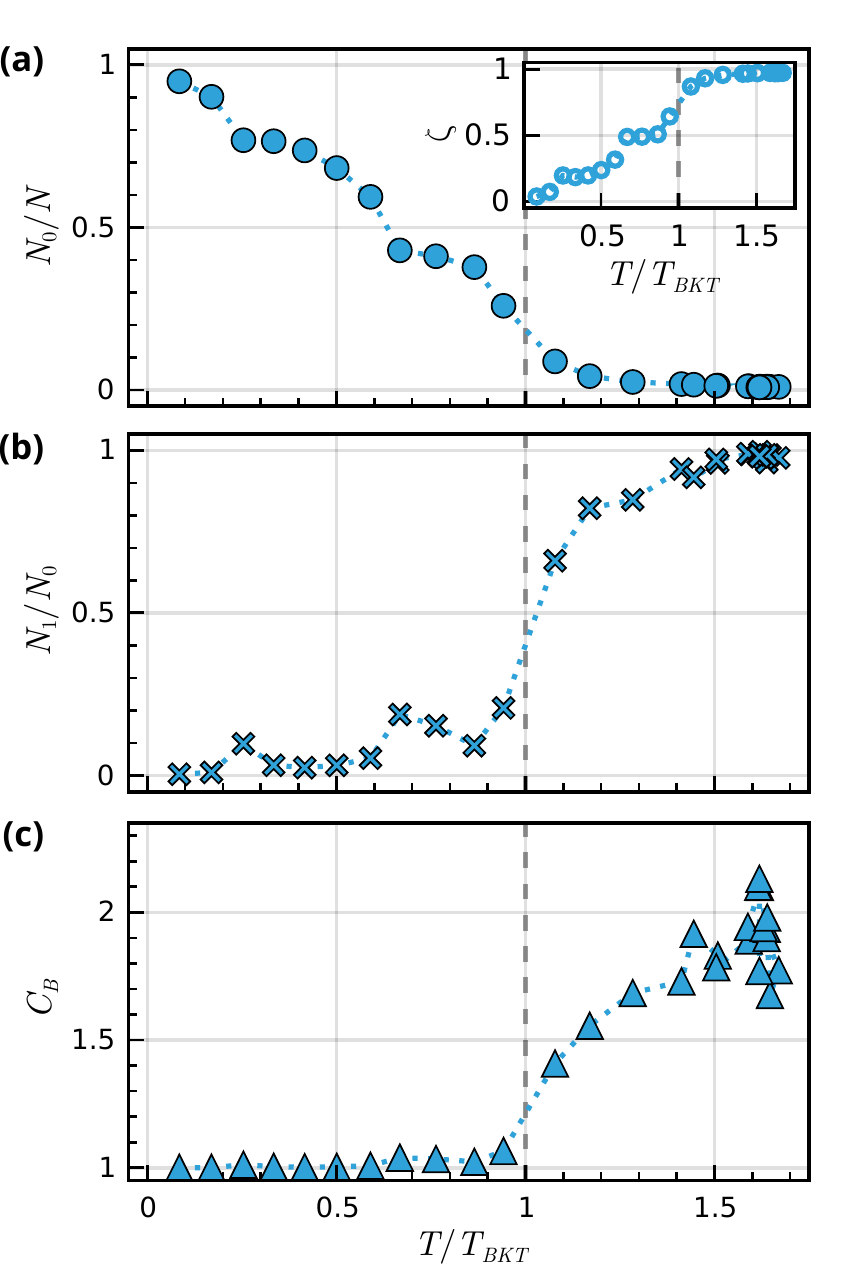}
    \caption{\protect Results for $\Lambda=1$ (2D limit). (a) Equilibrium condensate fraction $N_{0}/N$ as a function of rescaled temperature. The inset plots the quantity $\zeta=(N_{\rm QC}-N_{0})/N_{\rm QC}$, which highlights the difference between quasicondensate and condensate density as a function of rescaled temperature. (b) Equilibrium ratio of lowest modes $N_1$ and $N_0$ as a function of rescaled temperature. (c) Equilibrium Binder cumulant $C_{B}$ as calculated from the Penrose-Onsager condensate as a function of rescaled temperature. 
    In each plot, temperature has been rescaled to the BKT critical temperature (\ref{eq:tbkt}) for an infinite 2D system and marked as a vertical dashed line. }
    \label{fig:BKTplot}
\end{figure}
\begin{figure}[!ht]
    \centering
    \includegraphics[width=0.95\linewidth,keepaspectratio]{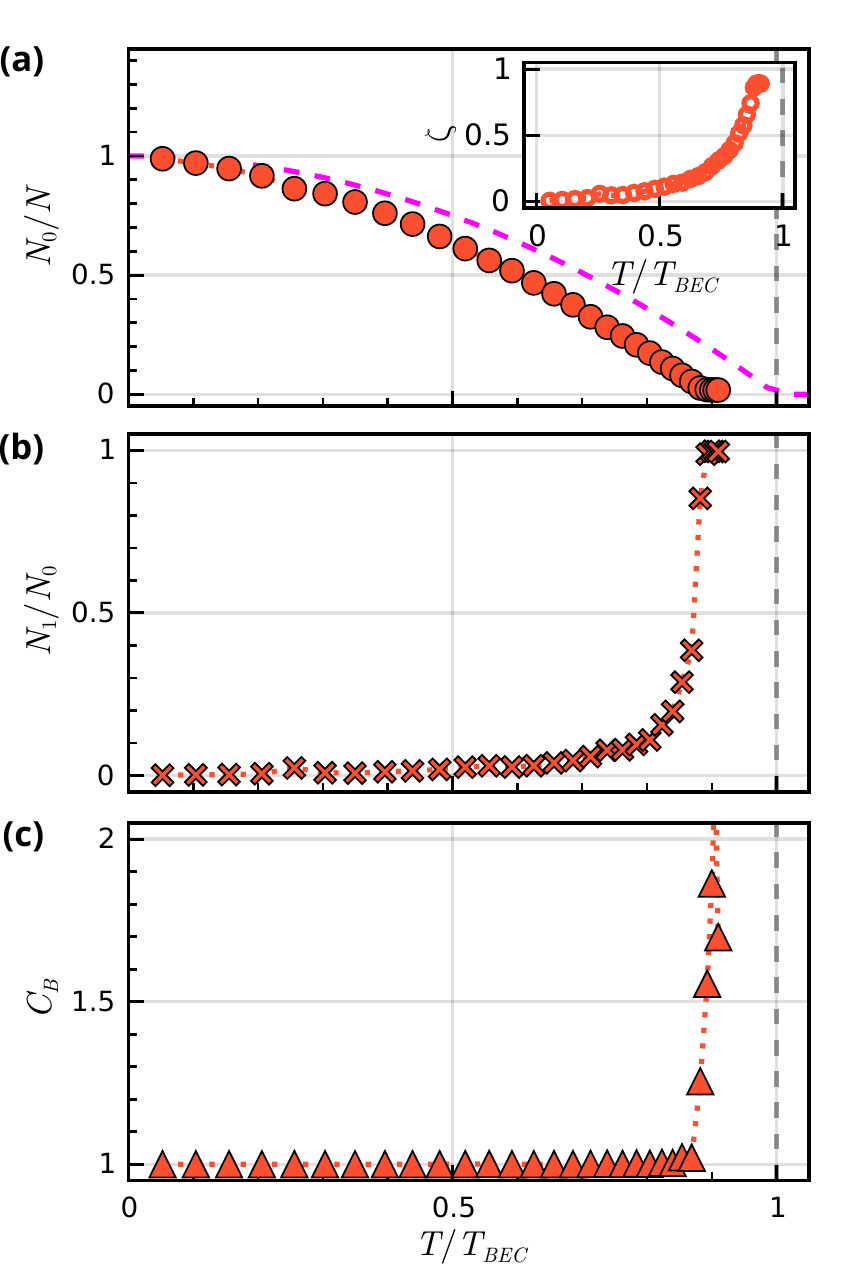}
    \caption{\protect Results for $\Lambda=50$ (3D limit). (a) Condensate fraction $N_{0}/N$, alongside the ideal gas law result (dashed magenta line). The inset plots the quantity $\zeta=(N_{\rm QC}-N_{0})/N_{\rm QC}$, which highlights the difference between quasicondensate and condensate density.  (b) Ratio of lowest modes $N_1$ and $N_0$. (c) Binder cumulant $C_{B}$. In each plot, temperature has been rescaled to the ideal gas BEC transition temperature defined in Eq.~(\ref{eqn:bec}) and marked as a vertical dashed line.}
    \label{fig:BECplot}
\end{figure}

To characterize the 2D phase transition in our numerics, we plot
in Fig.~\ref{fig:BKTplot} the relevant quantities discussed above (namely $N_0/N$, $N_1/N_0$, $C_B$ and $\zeta$) for $\Lambda = 1$ as a function of scaled temperature, $T/T_{\rm BKT}$. Here $T$ corresponds to the temperature of the thermal reservoir in contact with which the Bose gas has reached equilibrium, and $T_{\rm BKT}$ is defined in Eq.~(\ref{eq:tbkt}). Since the transition is not sharp in a finite size system~\cite{Foster10Vortex,Comaron19Quench,Gawryluk19Signatures} the determination of the actual transition temperature depends on the criterion and the indicator that is chosen to define it. The sharpest indicator is the Binder cumulant [Fig.~\ref{fig:BKTplot}(c)], which suggests that the critical temperature in the trapped gas is indeed very close to $T_{\rm BKT}$, i.e., the expected value in the thermodynamic limit. The occupation of the lowest states is smooth across the BKT transition [Fig.~\ref{fig:BKTplot}(a)-(b)], with the quasicondensate density remaining significantly larger than the condensate fraction 
$N_0/N$: the latter is evident by the fact that $\zeta$ remains finite and significantly nonzero for all $T<T_{\rm BKT}$. Overall, the results that we obtain here with a 3D SPGPE in a tight transverse confinement ($\Lambda=1$) are fully consistent with the purely 2D SPGPE simulations of Ref.~\cite{Comaron19Quench}, where the relevant features of the transition were discussed in detail. This validates our 3D formulation of what is essentially 2D physics within the hybrid-basis approach we enact.

We note on passing that previous works have also used the average number of vortices at equilibrium as a signature of the BKT transition in the 2D gas (see, e.g.~\cite{Foster10Vortex,Comaron19Quench}); whilst we omit the results here (for consistency of presentation with the crossover and 3D case where such meaning is not well defined), we note that by employing a plaquette method to identify vortex cores across the central plane we indeed achieved, as expected, results which are in agreement with the purely 2D simulations of Ref.~\cite{Comaron19Quench}.

\subsection{\label{sec:3DLimit}3D limit: BEC transition}

The corresponding 3D limit can be well probed by our largest choice of $\Lambda=50$, as can be seen by comparing such results with the predictions for BEC in 3D. In particular, using the density of states 
\begin{equation}
g(\epsilon) =  \frac{mL_xL_y \epsilon}{2\pi \hbar^3  \omega_z}   
\end{equation}
with $\omega_z=\omega_{\rm ref}/\Lambda$, one can calculate the temperature at which the condensate forms in an  ideal gas in the same hydrid trap: 
\begin{equation}
T_\mathrm{BEC} = \sqrt{\frac{12\hbar^3\omega_{\rm ref}N}{\pi k_B^2m\Lambda L_xL_y}}.
\label{eqn:bec}
\end{equation}
Due to interaction and finite-size effects, the actual transition temperature of a confined weakly-interacting Bose gas is expected to be downwardly shifted \cite{Pitaevskii16Bose,Dalfovo99Theory,Giorgini96Condensate,Davis06Critical,Liu20Kibble}, the shift depending on the type of confinement. 

As in the 2D case, we calculate the Binder cumulant, the fractions $N_0/N$ and $N_1/N_0$, and the quasicondensate density. For each equilibrium configuration at a temperature $T$ we calculate the total number of atoms, $N$, and we use it to estimate the ideal gas critical temperature $T_{\mathrm{BEC}}$ from Eq.~(\ref{eqn:bec}); then, all quantities are plotted as a function of the rescaled temperature $T/T_{\mathrm{BEC}}$. The results are shown in Fig.~\ref{fig:BECplot}. Compared to the 2D case, the equilibrium statistics exhibit a narrower critical region, with sharp transitions present in the ratio of dominant modes $N_1/N_0$ and the Binder cumulant. The condensate fraction  $N_0/N$ vanishes at the transition and its temperature dependence is similar to the one predicted for an ideal gas in the same trap  except for an downward shift. The observed transition temperature is at about $0.9T_{\mathrm{BEC}}$.

\begin{figure}[!ht]
    \centering
    \includegraphics[width=0.95\linewidth,keepaspectratio]{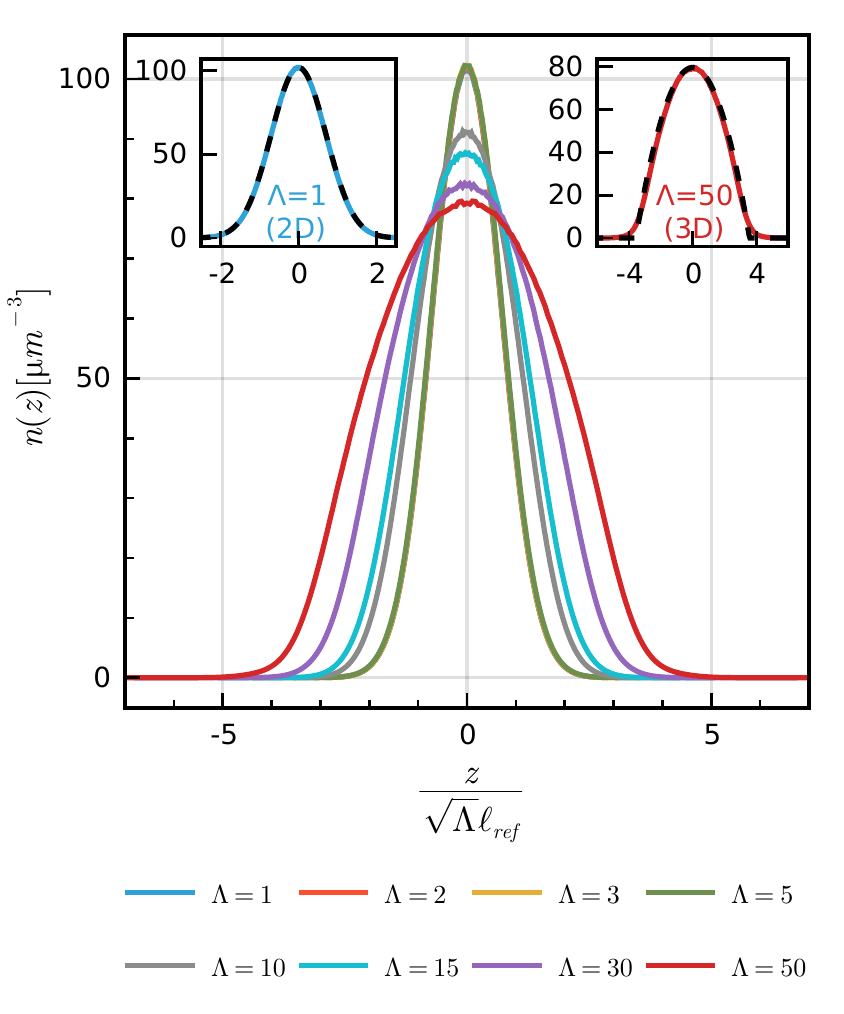}
    \caption{\protect Transverse profile of the total density of the gas for varying dimensionality $\Lambda$ at a fixed temperature $T=50$~nK. The coordinate $z$ is normalized to the harmonic oscillator length $l_z$. Note that the curves for $\Lambda=1,2,3$ are not visible, as they coincide with the one for $\Lambda=5$. Insets: zoomed-in profiles for extremes of $\Lambda=1$ (left) and $\Lambda=50$ (right) where dashed lines correspond to expected Gaussian and Thomas-Fermi ground-state profiles respectively.  }
    \label{fig:profiles}
\end{figure}

\subsection{\label{sec:Quasi2D}From 2D to 3D}

We can now explore the dimensionality crossover from 2D to 3D by varying $\Lambda$ . The shape of the density in the transverse direction directly witnesses the change of confinement. 
In Fig.~\ref{fig:profiles} we show the total density along the harmonically trapped $z$ direction rescaled to the corresponding harmonic oscillator length $l_z=\sqrt{\Lambda}l_{\rm ref}$ where we have sampled the averaged central points along $x$ and $y$.
All curves are the results of SPGPE simulations at $T=50$~nK, averaging over $\mathcal{N}\ge50$ equilibrium configurations. The two insets show the density in the two limiting cases $\Lambda=1$ (left) and $\Lambda=50$ (right), normalized to its central value at $z=0$. For $\Lambda=1$ the density profile is indistinguishable from the Gaussian of width $l_z$ (shown by the dashed blue line), which is the prediction for the lowest state of the ideal gas in the same harmonic trap. Indeed, in this limit one has $\hbar \omega_z \gg \mu$, 
and dynamics along the transverse direction become ``frozen" as atoms are locked into the ground state mode. In this regime, one can decouple the wavefunction into 
$\psi(\vec{r},t) = \psi(x,y,t)\phi_{z}(z)$, 
where the $z$ component coincides with the ground harmonic oscillator state; hence, after averaging the c-field in the planar directions, the density is expected to be  
$n_{\rm HO}(z) =(\pi \ell_z^2)^{-1/2} \exp (-z^{2} / \ell_{z}^{2})$. 

\begin{figure*}[!ht]
    \centering
    \includegraphics[width=0.65\linewidth,keepaspectratio]{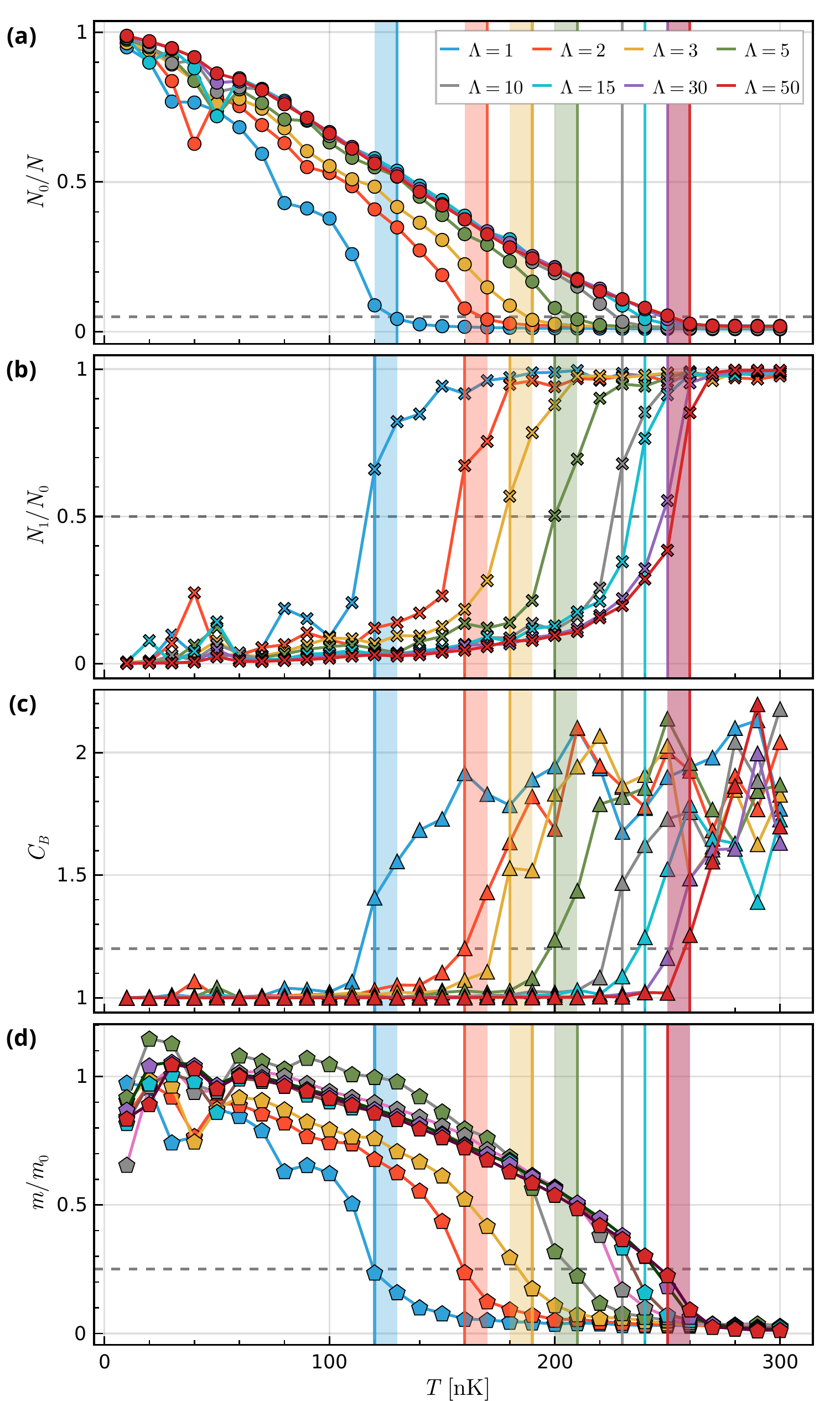}
    \caption{\protect 
Dependence of equilibrium parameters characterizing the degree of phase coherence of the system as a function of absolute temperature, plotted for different values of the dimensionality parameter $\Lambda$.
(a) Equilibrium condensate fraction $N_{0}/N$. 
(b) Ratio of lowest modes $N_1$ and $N_0$. 
(c) Binder cumulant $C_{B}$. 
(d) Order parameter $m$ normalised to the zero-temperature result $m_0$. 
The dashed horizontal line in each subplot corresponds to a cutoff value used to identify the transition for the respective equilibrium observable (with details of such choice discussed in Appendix~\ref{App:crit}). Each colour corresponds to a different value of dimensionality $\Lambda$ as indicated in legend. Solid vertical coloured lines in each subplot correspond to the first numerical point deemed to have crossed the phase transition for the respective quantity towards the incoherent regime, thus marking the identification of the critical temperature for each value of $\Lambda$ based on that physical quantity. Vertical coloured bands (where present) are identical throughout (a)-(d) and indicate the full range of numerically-identified critical temperature values across the different quantities plotted in (a)-(d). The midpoint of such band is subsequently chosen as our numerically identified critical temperature for each value of $\Lambda$.
}
\label{fig:megaplot}
\end{figure*}

In the opposite limit of large $\Lambda$, and for temperature much lower than the BEC transition temperature, the density is very well approximated by the inverted parabola predicted by the Thomas-Fermi approximation for a trapped BEC at $T=0$ in 3D, i.e.~the regime in which the kinetic energy term in the Gross-Pitaevskii equation can be neglected due to the large contribution from the nonlinear interaction term \cite{Dalfovo99Theory}, such that one can write the density as $n_{\textrm{TF}}(z)= (1/g) [\mu-V(z)]$ for  $V(z)<\mu$, and zero otherwise. This prediction corresponds to the dashed red line in the top right inset of Fig.~\ref{fig:profiles}.  In between the two limits the density profile changes smoothly from a Gaussian to an inverted parabola.

\begin{figure*}[!ht]
    \centering
    \includegraphics[width=0.99\linewidth,keepaspectratio]{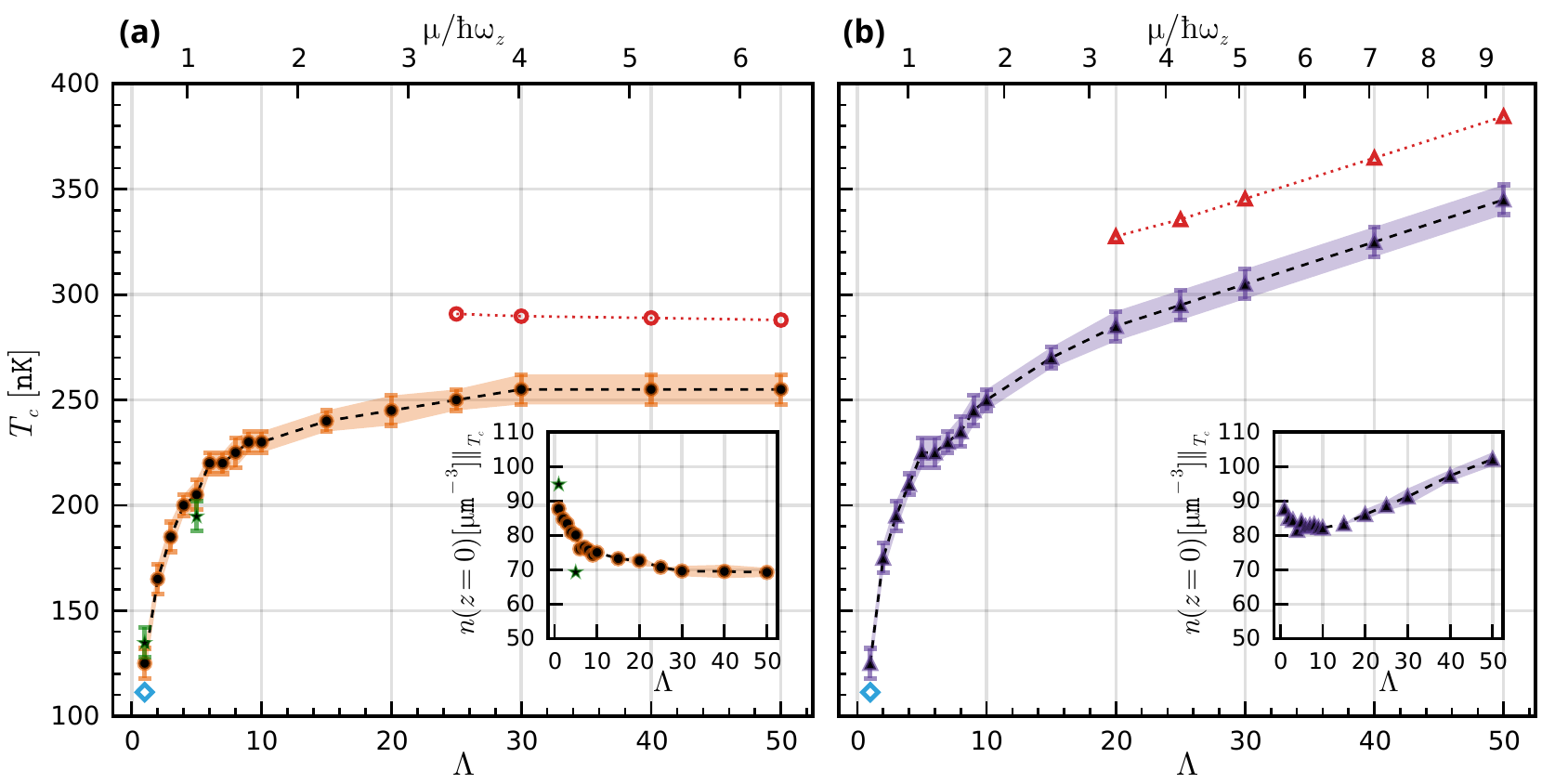}
    \caption{\protect 
    Numerically-extracted phase transition temperature $T_c$ (main plots), and central densities (insets) as a function of dimensionality parameter $\Lambda$, for two different protocols for our choice of chemical potential $\mu$ based on  Eq.~(\ref{eq:inputmu}). Specifically, they correspond to (see text):
    (a) a constant $\mu_{\rm 2D} = (4.64 \mu {\rm m}^{-2}) \hbar^2/m$ value, and (b) a value linearly interpolated between such $\mu_{\rm 2D}$ value, for  $\Lambda=1$, and $(3/2)\mu_{\rm 2D}$, for our most 3D case with $\Lambda=50$. 
    In both cases, black filled points correspond to the mean extracted transition temperature as determined by the equilibrium statistics (Fig.~\ref{fig:megaplot}), with dashed black lines as a guide to the eye.
    Error bars and shaded regions highlight the
    associated uncertainty, arising from a combination of the width of the bands in Fig.~\ref{fig:megaplot}
    and a systematic uncertainty of $\pm5\mathrm{nK}$ due to our limited resolution in probing distinct temperatures.
    Hollow blue diamonds mark the analytical $T_{\mathrm{BKT}}$ transition temperature at $\Lambda=1$, from Eq.~(\ref{eq:tbkt}).
    Red hollow points for $\mu / \hbar \omega_z > 3$, with dotted red lines as a guide to the eye, indicate the analytical 3D ideal gas temperature, $T_{\mathrm{BEC}}$, in our chosen geometry, using the same total atom number $N$ in Eq.~(\ref{eqn:bec}) as in the corresponding SPGPE simulation for the same $\Lambda$.
    Green points in (a) correspond to further simulations with two different values of $\mu$ at small $\Lambda$ aimed at demonstrating the limited sensitivity of the rapid initial growth of $T_c$ with $\Lambda$ on density.
    Insets: Total density at the center of the trap; for each $\Lambda$, the point represent an average of the central density in a few SPGPE simulations for $T\sim T_c$, and the coloured bands indicate the corresponding numerical uncertainty.
    }
    \label{fig:fig6}
\end{figure*}

 In Fig.~\ref{fig:megaplot} we plot the quantities $N_0/N$, $N_1/N_0$, and the Binder cumulant $C_B$, as a function of absolute temperature for a characteristic subset of the dimensionalities, namely $\Lambda=\{1,2,3,5,10,15,30,50\}$. Here, in addition, we also use the order parameter defined by \cite{Kobayashi16Quench,Comaron19Quench,Liu20Kibble}
\begin{equation}
    m=\frac{\left\langle \left|\int d\mathbf{r}~\Psi(\mb{r})\right|\right\rangle_\mathcal{N}}{\sqrt{\left\langle\int d\mathbf{r}~\left|\Psi(\mb{r})\right|^2\right\rangle_\mathcal{N}}} 
    \label{eqn:orderparam}
\end{equation}
which acts as a further measure of the degree of degeneracy of the system. 
It is worth mentioning that the SPGPE is inherently a high-temperature theoretical framework, in the sense that the equilibration processes described by the SPGPE become inefficient when $T/T_c$ is significantly less than $1$. At such low temperatures, the criterion that the thermal reservoir contains many weakly populated thermal modes is not met \cite{Blakie08Dynamics} and large fluctuations in equilibrium statistics are expected. The order parameter $m$ is particularly  susceptible to such fluctuations, as can be in Fig.~\ref{fig:megaplot}(d) for the points below about $50$~nK.

To estimate the transition temperature $T_c$ we select, for each probed quantity, a relevant cut-off value marking the transition from coherent to incoherent regime, with such cut-off value indicated by the horizontal dashed line in each panel.
Using this as our critical value, we identify a specific temperature at which the transition likely occurs, as the first value within the incoherent region which crosses the indicated threshold.
Although in some cases such numerical temperature identification  perfectly overlaps (within our temperature and numerical resolution) across all panels (see $\Lambda =15,\,30$ cases), in most cases the different probed quantities yield slightly different numerical values for the critical temperature.
Such an effect is accommodated  by adding to each plot a coloured vertical band, indicating the uncertainty in such identification across all probed markers for a given value of $\Lambda$.
The value of $T_{\rm c}$ is then identified as the midpoint of such bands, with an associated uncertainty given by half the width, combined with a systematic uncertainty of $\pm 5$nK due to the discrete range of probed temperatures (further details of the numerical determination of $T_{\rm c}$ are given in Appendix.~\ref{App:crit}).

Such identification of the critical temperature is then shown as a function of the dimensionality parameter $\Lambda$ in Fig.~\ref{fig:fig6}, with panels (a) and (b) based on two different numerically-generated datasets, differing in the way $\mu$ is changed when transitioning from 2D to 3D. Fig.~\ref{fig:fig6}(a) corresponds to results extracted from the data shown in Fig.~\ref{fig:megaplot}, for which the chemical potential $\mu$ is defined according to Eq.~(\ref{eq:inputmu}) with $\mu_{\rm 2D}$ independent of $\Lambda$. 
Fig.~\ref{fig:fig6}(b) is based on a different protocol for our chemical potential choice, where we substitute the first term on the right-hand-side of Eq.~(\ref{eq:inputmu}) with a function of $\Lambda$ that linearly interpolates from its quoted 2D value, $\mu_{\rm 2D}$ for $\Lambda=1$, towards its 3D value $(3/2) \mu_{\mathrm{2D}}$, assumed here to arise when $\Lambda=50$. This choice is designed to match the $T=0$ Bogoliubov sound speed across the dimensional extremes (see Sec.~23.1 in \cite{Pitaevskii16Bose}). The corresponding ratios of $\mu / \hbar \omega_z$ differ across the two panels, as shown by the upper labels.

In both cases the numerically identified critical temperature exhibits a very rapid increase with increasing value of $\Lambda$ between the expected limits.
For small $\Lambda$ we recover the analytically-expected BKT transition temperature, given by Eq.~(\ref{eq:tbkt}) and marked by the leftmost hollow blue diamond.
We observe a monotonic increase of $T_{\rm c}$ with $\Lambda$ as the system transitions to 3D, with the dominant increase in the critical temperature occurring for $\mu / \hbar \omega_z \lesssim 2$.
As $\mu/\hbar \omega_z$ increases beyond that, the critical temperature dependence on $\Lambda$ rapidly mimics the one expected analytically for an ideal Bose gas in the 3D limit (Eq.~(\ref{eqn:bec})) for the given atom number
(strictly valid in the $\Lambda \rightarrow \infty$ limit): these are marked, for $\mu/\hbar \omega_z > 3$, by the hollow red symbols in each subplot; note that the increase seen in panel (b) for large $\Lambda$ is a direct consequence of our chemical potential protocol, associated with a linearly increasing density/atom number with $\Lambda$. 
As expected, our numerically-extracted values are consistently lower than the analytical ideal gas ones, due to finite-size and interactions effects.

The insets plot the dependence of the central condensate density, averaged along the planar directions and evaluated at the phase transition $T_c$ for each given geometry, as a function of $\Lambda$, with the different observed behaviour arising from the different chemical potential protocols adopted. Although the densities exhibit up to $20\%$ variation, it is important to highlight that the dominant dependence of $T_{\rm c}$ on $\Lambda$ for small values of $\Lambda$ is not a consequence of the changing density. To this aim, 
we have added two further {\em ad hoc} simulation points, spanning the density extrema in (a) and observe no noticeable change (within our uncertainties) to the numerically-extracted value for $T_{\rm c}$.
We thus directly conclude that the observed changes to the location of the phase transition are a consequence of dimensionality. 

This above observation reveals that a marginally stronger harmonic trapping strength can lead to large differences in phase transition temperature when exploring BKT physics. This is an important consideration in view of the current experiments with quasi-2D configurations. It is also worth noticing that the range $\mu/\hbar \omega_z$ where $T_c$ ramps up toward the asymptotic 3D behavior is the same where the healing length, $\xi$, becomes comparable to the transverse size of the gas, $\ell_z$; in particular, for $\mu/\hbar \omega_z=2$ the healing length at $T=0$ is $\xi=\hbar/\sqrt{2mgn(0)} \sim (1/2) \ell_z$.  In terms of vortices, as mentioned earlier when commenting Fig.~1, this implies a transition from point-like defects in a 2D background to vortical filaments in a 3D superfluid. This is not surprising, but our simulations provide a quantitative and systematic description of the dimensional crossover in terms of equilibrium properties, going beyond the known qualitative picture. In addition, Fig.~\ref{fig:fig6} also suggests an interpretation of the behaviour of $T_c$ in terms of BEC vs. BKT physics. In the large $\Lambda$ limit, if one cools down the Bose gas close to $T_c$, the quantum degeneracy manifests itself almost at the same temperature at which the lowest state becomes macroscopically occupied and the BEC soon forms together with the superfluid phase. Conversely, for small $\Lambda$, due to stronger effects of fluctuations, quantum degeneracy is associated to a quasicondensation, where many modes have large occupation; one then needs to cool the gas further in order to allow the system to develop coherence and superfluidity as a result of vortex-pair binding. For gases with similar densities, as in our simulations, this implies a critical temperature lower in 2D than in 3D. 

\section{\label{sec:conclusion} Conclusions}
We have performed a systematic analysis of the effects of temperature and dimensionality in the ultracold Bose gas trapped via the box-harmonic hybrid potential, employed in current experiments. We have demonstrated a continuous shift in the phase transition temperature of the trapped Bose gas as a function of dimensionality, for consistent atomic densities, with the fastest change occurring for $\mu \lesssim 2 \hbar \omega_z$. Our work highlights the need for great care in choice of experimental trapping frequencies to ensure true ground state occupation in the lowest transverse modes whilst exploring BKT physics. Similarly, we highlight interesting properties in the thermality of dimensionally reduced Bose gases, relevant for experimental observation Bose-Einstein condensation in quasi-dimensional trapping geometries \cite{Choi13Observation,Hadzibabic06Berezinskii,Neely10Observation,Gauthier19Giant,Johnstone19Evolution}. This work motivates further research and insights on questions related, for instance, to vortex topology in response to dimensional quenches and to the nature of linear and nonlinear collective excitations of the gas in the dimensional crossover.

\section*{Data availability}
Data supporting this publication is openly available under an
“Open Data Commons Open Database License”. Additional metadata are available 
at: \href{URL}{TBA}

\begin{acknowledgments}
This work has benefited from Q@TN, the joint lab between University of Trento, FBK-Fondazione Bruno Kessler, INFN - National Institute for Nuclear Physics and CNR - National Research Council. FD and NAK also acknowledge the support of Provincia Autonoma di Trento through the agreement with the INO-CNR BEC Center.
We acknowledge financial support from the Quantera ERA-NET cofund
project NAQUAS through the Consiglio Nazionale delle
Ricerche (FD) and the Engineering and Physical Science Research Council, Grant No. EP/R043434/1
(IKL, NPP).

\end{acknowledgments}

\appendix

\section{\label{App:Numerics}Numerical method}

To solve the SPGPE for the considered box-harmonic hybrid potential, Eq.~(\ref{eqn:potential}), we adopt a hybrid basis composed of plane waves along the $x$ and $y$ axes and the harmonic oscillator basis along the $z$ axis. Since the SPGPE simulations are time consuming, we prefer to make the calculations faster by implementing efficient Fourier and  inverse  Fourier  transformations in the $xy$-plane. This requires the use of periodic boundary conditions. We thus embed the physical $L_x \times L_y$ box into a slightly larger auxiliary ${\cal L}_x \times {\cal L}_y$ box. The potential outside the physical box is taken to be very large (decades larger than the chemical potential) so that the density is negligible in that region.  Periodic boundary conditions are then applied to the auxiliary box. 

The SPGPE is numerically solved in its dimensionless form in which the physical quantities and variables are scaled by reference length, energy and times scales, $l_\mathrm{ref}$, $\hbar\omega_\mathrm{ref}$ and $\omega_\mathrm{ref}^{-1}$, respectively, according to their dimensions, and in the following the dimensionless quantities and variables are denoted by prime notation. The c-field is expanded in the hybrid basis
\begin{equation}
    \Psi'(\mathbf{r}',t')=\sum_{\varepsilon_{pqn}\leq\epsilon'_\mathrm{cut}}A_{pqn}'(t')\phi_p(x')\phi'_q(y')\varphi_n'(z')
\end{equation}
where the dimensionless single-particle energies are 
\begin{equation}
    \epsilon_{pqn}'=2\pi^2 \left[\frac{ p^2}{\mathcal{L}_x'^2}+\frac{ q^2}{\mathcal{L}_y'^2}\right]+\frac{1}{\Lambda}\left(n+\frac{1}{2}\right)  \ .
\end{equation}
The wavefunctions are 
\begin{eqnarray}
\phi_p'(x') &=& (1/\sqrt{\mathcal{L}_x'})\exp(i2\pi p x'/\mathcal{L}_x') \\ 
\phi_q'(y') &=& (1/\sqrt{\mathcal{L}_y'})\exp(i2\pi q y'/\mathcal{L}_y') \\
\varphi_n'(z') &=& \frac{\Lambda^{1/4}}{\sqrt{2^nn!}\,\pi^{1/4}}H_n\left(\sqrt{\Lambda}z'\right)e^{-\Lambda z'^2/2 }
\end{eqnarray}
where $H_n$ are the Hermite polynomials and $p,q,n$ are integer quantum numbers with $p,q\ge 1$ and $n\ge 0$. This hybrid basis satisfies the eigenequation 
\begin{equation}\begin{array}{l}
\displaystyle    \left[-\frac{\nabla'^2}{2}+\frac{1}{2\Lambda^2}z'^2\right]\phi_p(x')\phi'_q(y')\varphi_n'(z')\\\\\displaystyle\quad=\varepsilon'_{pqn}\phi_p(x')\phi'_q(y')\varphi_n'(z')
\end{array}\end{equation}
and forms a complete set obeying the orthogonality condition 
\begin{equation}
\int d\mathbf{r}\phi_{p'}^\ast(x)\phi_{q'}^\ast(y)\varphi_{n'}^\ast(z)\phi_p(x)\phi_q(y)\varphi_n(z)=\delta_{pp'}\delta_{qq'}\delta_{nn'} \ .
\end{equation}
The amplitude of each mode in the c-field can be expressed in the form
\begin{equation}\begin{array}{rl}
A_{pqn}'(t')=&\displaystyle\int d\mathbf{r}'\phi_{p}'^\ast(x)\phi_{q}'^\ast(y)\varphi_n'^\ast(z)\Psi'(\mathbf{r}',t')
\\\\
\equiv&\mathcal{F}_{x,p}\bigg[\mathcal{F}_{y,q}\Big[\mathcal{H}_{z,n}\big[\Psi'(\mathbf{r}',t')\big]\Big]\bigg]
\end{array}\end{equation}
where $\mathcal{F}_{x,p}$ and $\mathcal{F}_{y,q}$ denote the Fourier transform and $\mathcal{H}_{z,n}$ is the Hermite transformation. The corresponding inverse transformation is 
\begin{equation}\begin{array}{rl}
\Psi'(\mathbf{r}',t')=&\displaystyle\sum_{p,q,n}\mathcal{F}_{x,p}^{-1}\bigg[\mathcal{F}_{y,q}^{-1}\Big[\mathcal{H}_{z,n}^{-1}\big[A_{pqn}'(t')\big]\Big]\bigg].
\end{array}\end{equation}
The Fourier and inverse Fourier transformation can be straightforwardly computed by fast Fourier transformation with homogeneous grids along $x$ and $y$ axes.
The Hermite transformation can be computed with the Hermite-Gaussian quadrature,
a form of Gaussian quadrature for approximating the value of integrals by a summation with $n$ points, given by
\be
\int dx\ e^{-z'^2}f(z')dz'\approx\sum_{i=1}^{n_{max}}w_{i,n}\,f(\alpha_i) \ ,
\ee
where
\be
w_{i,n}=\frac{2^{n-1}n!\sqrt{\pi}}{n^2[H_{n_\mathrm{max}-1}(\alpha_i)]^2}
\ee
are roots of the Hermite polynomial $H_{n_\mathrm{max}} (\alpha_i)$, for $i=1,\;2\;...,\;n$~\cite{Blakie08Dynamics,Blakie08Numeric}.
In order to compute the Hermite transformation accurately, the grid is set by $\alpha_i$ and is not homogeneous along the $z$-axis. Then we can write
\begin{equation}
\mathcal{H}_{z,n}[\Psi'(\mathbf{r}',t')]=\sum_iw_{i,n}\frac{H_n(\alpha_i)}{\sqrt{\Lambda}}e^{\Lambda \alpha_i^2/2}\Psi(x',y',\alpha_i',t') \ ,
\end{equation}
and this can be numerically implemented by a product of a matrix with indices $n$ and $i$ times a vector with index $i$.
The inverse Hermite transformation is the product of the basis weight and the $n$-th harmonic basis,
\begin{equation}
\mathcal{H}_{z,n}^{-1}\left[A_{pqn}'(t')\right]=A_{pqn}'(t')\varphi'_n(z') \ .
\end{equation}
During the SPGPE evolution addressed below, the $z$-position grid is set by the roots of the Hermite polynomial with $n_\mathrm{max}=300$ in all our simulations.
Besides, it is worth noting that unlike the Hermite transformation, one can reconstruct the wave function with arbitrary spatial resolution from the the inverse transformation with the knowledge of $A_{{pqn}}$, and thus it will allow us to visualise the data with finer grid spacing in $z'$.

The equation of motion of the basis amplitude $A_{pqn}$ for $\varepsilon_{pqn}'\leq\varepsilon_\mathrm{cut}$ can be obtained by the SPGPE
\begin{eqnarray}
    i\frac{\partial}{\partial t'}A_{pqn}'(t') & = & (1-i\gamma)\left[\varepsilon_{pqn}'+\mathcal{P}[G_{pqn}']-\mu'\right]A_{pqn}'(t) \nonumber \\
    &+&\tilde{\eta}_{pqn}'(t')
\end{eqnarray}
where 
\begin{eqnarray}
G_{pqn}'(t') & = & \mathcal{F}_{x,p} \bigg[\mathcal{F}_{y,q} \Big[\mathcal{H}_{z,n} \big[V'(x',y') \nonumber \\
& + &  g'\left|\Psi'(\mathbf{r}',t'\right|^2\Psi'(\mathbf{r}',t')\big]\Big]\bigg]
\end{eqnarray}
is the combination of external box potential and the nonlinear term. Here, we introduced $g'=4\pi a_s/l_{\rm ref}$ as the dimensionless interaction strength. Meanwhile the complex white noise follows the correlation relation
\begin{equation}
    \left\langle\tilde{\eta}_{ijk}'^\ast (t^{\prime\prime})\tilde{\eta}_{pqr}'(t')\right\rangle=2\gamma \frac{k_BT}{\hbar\omega_\mathrm{ref}}\delta(t'-t'')\delta_{pi}\delta_{qj}\delta_{nk}
\end{equation}
only for modes below the cutoff. The projector limits the modes evolved in the dynamic for modes satisfying $\varepsilon_{pqn}\leq\varepsilon_\mathrm{cut}$ while higher modes could be created during the $s$-wave collision.
The time integral is solved by a Runge-Kutta $4^{\textrm{th}}$ order routine with a time step which varies across dimensionalities but is typically $dt\sim dxdy\Lambda/2$. Numerics and visualisation are performed using Julia \cite{Bezanson17Julia,Omlin20Solving,Danisch21Makie}, whereby we utilise CUDA to offload calculations to a graphical processing unit (GPU). This project utilised the Rocket High Performance Computing service at Newcastle University.

\section{\label{App:density} Determination of total density}

The SPGPE partitions atoms into a  highly-occupied c-field $\Psi(\vec{r},t)$ coupled to a reservoir of atoms in the incoherent states above the cutoff. The density of atoms in the c-field is simply $n_c(\mathbf{r},t)= |\Psi(\mathbf{r},t)|^2$ and their number is $N_c=\int d{\textbf{r}} |\Psi|^2$. Instead, for the atoms above the cutoff one can assume that they occupy the single particle states of an ideal gas in the same trap with mean occupation number at equilibrium given by the Bose-Einstein distribution~\cite{Rooney10Decay}. In this calculation the trap is the physical one, which has the $L_x \times L_y$ box with infinite hard walls. 
In order to avoid confusion with the previous Appendix, here we use a slightly different notation for the quantum numbers. The eigenfunctions for a particle in our hybrid trap are given by
\begin{equation}
\begin{aligned}
\Psi_{\bf n} (x,y,z) = & ~ c \sin (n_x \pi x/L_x ) \sin (n_y \pi y/L_y) \\& \times H_{n_z} (z/\ell_z)  \exp (-z^2/2 \ell_z^2)  \, ,
\end{aligned}
\end{equation}
where $c$ is a normalisation constant, fixing the norm of a given state to 1, the functions $H_{n_z}$ are the Hermite polynomial, and ${\bf n}=(n_x,n_y,n_z)$ are integer quantum numbers with $n_x$, $n_y > 1$ and $n_z \ge 0$. The eigenenergies are 
\begin{equation}
E = \frac{\pi^2 \hbar^2}{2m}  \left( \frac{n_x^2}{L_x^2} + \frac{n_y^2}{L_y^2} \right) + \left(  n_z + \frac{1}{2}  \right) \hbar \omega_z     \ .
\label{eq:enxnynz}
\end{equation}
Thus the density associated to atoms above the cutoff energy $\epsilon_{\rm cut}$, treated as an ideal gas, can be estimated as 
\begin{equation}
n_{\cal I} (x,y,z) =   \sum_{\tilde{E}_{\bf n} > \tilde{\epsilon}_{\rm cut} }  \frac{|\Psi_{\bf n} (x,y,z)|^2}{\exp [(\tilde{E}_{\bf n}-\tilde{\mu})/k_BT]-1 }    \ .
\end{equation}
For convenience, here we have removed the zero point energy
$(1/2)\hbar \omega_z$ from the energy, cutoff energy and chemical potential, by defining 
$\tilde{E}_{\bf n} = E -   \hbar \omega_z/2 $, $\tilde{\epsilon}_{\rm cut}=\epsilon_{\rm cut} -\hbar \omega_z/2$ and $\tilde{\mu}= \mu -\hbar \omega_z/2$. 
We have also ignored interference effects among different eigenfunctions in the expression of the density $n_{\cal I}$; this is consistent with the assumption that the states above the cutoff energy are incoherent and the effects of the relative phase vanish after configuration averages. The spatial integral of this density gives the incoherent atom number $N_{\cal I}$.

Now we use the fact that $k_BT$ is always much larger than the energy spacing between planar states with different $n_x$ and $n_y$, so that their spectrum forms a continuum and the corresponding sums can be replaced with an integral, with the 2D density of states $mL_xL_y/(2 \pi \hbar^2)$. In addition,  all $\sin^2$ functions sum up to an areal density along $x$ and $y$ which can be assumed to be constant except for negligible boundary effects. Under these assumptions the incoherent density becomes
\begin{equation}
\begin{aligned}
n_{\cal I} (z)  &= \frac{m}{2 \pi \hbar^2}  \sum_{n_z=0}^\infty  |\psi_{n_z} (z) |^2 
\\& \times \int_{E_{\rm min}}^\infty dE_{xy} \, \frac{1}{\exp [(E_{xy}+n_z\hbar\omega_z-\tilde{\mu})/k_BT]-1 } 
\, ,
\end{aligned}
\label{eqn:transversedens}
\end{equation} 
where $\tilde{E}_{\bf n} = E_{xy} + n_z  \hbar \omega_z$. The wavefunction is  
\begin{equation}
\psi_{n_z} (z) =  [ \sqrt{\pi} \ 2^{n_z} n_z! \ \ell_z ]^{-1/2}  H_{n_z} (z/\ell_z)  \exp (-z^2/2 \ell_z^2) 
\end{equation}  
and has norm $1$.
The minimum energy is defined as 
\begin{equation}
E_{\rm min} =
\begin{cases}
\tilde{\epsilon}_{\rm cut}-n_z\hbar \omega_z & {\rm for} ~ \tilde{\epsilon}_{\rm cut}>n_z\hbar \omega_z \\
0 & {\rm for} ~ \tilde{\epsilon}_{\rm cut} \le n_z\hbar \omega_z \ .
\end{cases}
\end{equation}

The atoms above the cutoff must be in thermal equilibrium with those in the c-field and should share the same chemical potential. This implies that $\tilde{\mu}=\mu_{\rm 2D}$, given in Eq.~(\ref{eq:inputmu}). Moreover, the cutoff is fixed by $\tilde{\epsilon}_{\rm cut}=\mu_{\rm 2D}+k_BT \ln 2$.  
Using these expressions, the integral of (\ref{eqn:transversedens}) can be solved analytically, allowing one to derive the expression
\begin{equation}
\begin{aligned}
n_{\cal I} (z) =  \frac{ T'}{2\pi \ell_{\rm ref}^2 }  \bigg( & \sum_{n_z=0}^{\bar{n}_z-1}  |\psi_{n_z} (z) |^2  \ln 2 ~- \\&
\sum_{n_z=\bar{n}_z}^{\infty}  |\psi_{n_z} (z) |^2 \ln \left( 1- e^{\phi_{n_z}}\right)  \bigg)  \, .
\end{aligned}
\label{eqn:transverse_therm_dens}
\end{equation}
The dimensionless temperature is defined as $T'=T/T_{\rm ref}$, with reference temperature $T_{\rm ref}= \hbar \omega_{\rm ref}/k_B$. In addition, we have introduced the constant $a =  \mu_{\rm 2D}/(\hbar \omega_{\rm ref})$ and the expressions 
$\phi_{n_z}  = \left( a - n_z/\Lambda \right)/T'$ and $\bar{n}_z  = \lceil  \left( a + T' \ln 2 \right) \Lambda \rceil$. The spatial integral of the density gives the incoherent atom number
\begin{equation}
N_{\cal I} = \frac{L_xL_y}{2\pi \ell_{\rm ref}^2 } \Lambda T' ( a \ln 2   +   b T' ),  \ 
\end{equation} 
where $b = \int_{\ln2}^\infty dt~t/(e^t-1) = 1.0592$.
These expressions allow us to calculate the total density $n(z)=n_c(z)+n_{\cal I} (z)$ and the total number of atoms $N=N_c+N_{\cal I}$. 

\begin{figure}
    \centering
    \includegraphics[width=0.99\linewidth,keepaspectratio]{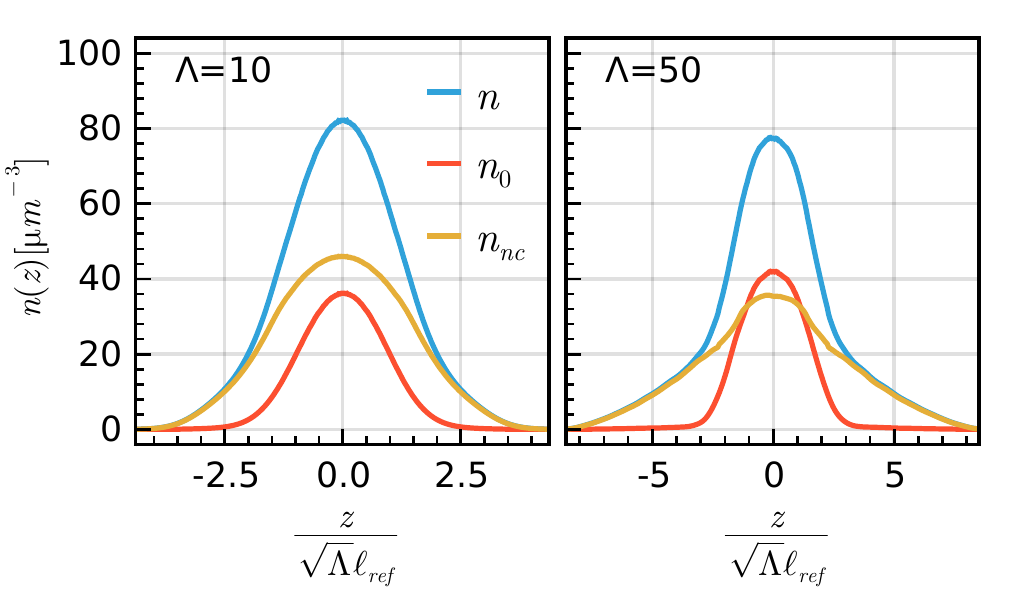}
    \caption{\protect Total ($n$), condensate ($n_0$) and non-condensate ($n_{nc} = n - n_0$) density profiles along the transverse direction plotted for a quasi-2D ($\Lambda=10$) and 3D ($\Lambda=50$) trapping geometry at temperature T=200$nK$, for which $T/T_c = 0.83$ and $T/T_c= 0.75$ respectively.
    }
    \label{fig:fig7}
\end{figure}

It is finally worth noticing that $n_{\cal I}$ must not be confused with the ``thermal" (or non-condensate) gas density as usually defined in 3D Bose gases when a condensate is present. The non-condensate density $n_{nc}(z)$ can be estimated as the total density $n(z)$ minus the density $n_0(z)$ of the atoms in the condensate, i.e., the eigenstate with the largest eigenvalue in the  Penrose-Onsager diagonalisation of the one-body density matrix. Such a thermal density takes contributions from both the c-field below and the incoherent states above the cutoff. Two examples are shown in Fig.~\ref{fig:fig7} at temperatures close to the phase transition for two different values of $\Lambda$.

\section{\label{App:crit}Identification of the critical region}
Localisation of the phase transition and the respective critical region is performed by numerical thresholding for each considered equilibrium parameter as displayed in Fig.~\ref{fig:megaplot}. To characterise the transition we find both the minimum and maximum temperature at which our selection of equilibrium statistics indicate a crossing
from coherent to incoherent behaviour, using the first identified incoherent point for each equilibrium parameter, and then averaging across the identified extrema.
Then, accounting for the thermal resolution of our simulations, we combine the half-width of the numerically-extracted band (where present) with an additional independent uncertainty of $\pm5\mathrm{nK}$
to define the overall width of our critical region.

We emulate previous works~\cite{Liu20Kibble} and stipulate the crossing of a phase transition occurs once the condensate fraction falls below $5\%$ and introduce the cutoff 
$(N_{0}/N)^{\textrm{cut}}=0.05$ to signpost a crossing of the critical region. At the phase transition point, occupation levels in the lowest and second-lowest mode should become comparable. Considering the ratio of the two lowest system modes $r_{\Xi} = N_{1}/N_{0}$, one would expect an imminent transition when there are half as many atoms in the second-lying mode as in the first. As such, we introduce a cutoff value $r_{\Xi}^{\textrm{cut}}=0.5$. The Binder cumulant, as previously defined in Eq.~(\ref{eqn:binder}), is a well established parameter to signal the crossing of a phase transition \cite{Kobayashi16Quench,Comaron19Quench}. In the thermodynamic limit, the critical value is known to be $C^{\infty}_{B}=1.2430$ \cite{Campostrini01Critical}, whereas for trapped systems, where finite-size effects can manifest, this value is lower \cite{Bezett09Critical}. In this vein, we select a critical value of $C^{\textrm{cut}}_{B}=1.2$ to indicate a crossing of the critical region. Lastly, we consider the order parameter $m$ as defined in Eq.~(\ref{eqn:orderparam}) normalised to its value at zero temperature $m_{0}$. For a zero-temperature system $m/{m_0}=1$, falling to $m/{m_0}\sim 0$ across the transition \cite{Kobayashi16Quench,Larcher18Dynamical}. To capture the transition, we introduce a threshold value of $m^{\mathrm{cut}}=0.25$. Using each of the aforementioned threshold values in combination, a systematic simultaneous crossing of the  critical region is revealed for each of the equilibrium statistics chosen. This allows us to measure the phase transition as a function of the system dimensionality, as showcased in Fig.~\ref{fig:fig6}. 

\bibliography{main.bib}

\end{document}